\documentclass[a4paper,10pt]{article}
\usepackage{graphicx}
\usepackage{dcolumn}
\usepackage{bm}
\usepackage{misccorr}

\long\def\symbolfootnote[#1]#2{\begingroup%
\def\thefootnote{\fnsymbol{footnote}}\footnote[#1]{#2}\endgroup}

\begin{document}

\textrm{ \begin{flushleft} \Large \bf A hybrid approach for predicting the distribution of 
vibro-acoustic energy in complex built-up structures \end{flushleft}  }

\textrm{ \begin{flushright} { \large \bf Dmitrii N. Maksimov\symbolfootnote[1]{mdn@tnp.krasn.ru}  and Gregor Tanner} \\
{ \small School of Matematical Sciencies, } \\
{ \small the University of Nottingham,  NG7 2RD} \\
{ \small  United Kingdom}
\end{flushright}  }

\

{ \noindent \it Finding the distribution of vibro-acoustic energy in complex built-up structures in the mid-to-high frequency regime is a difficult task. In particular, structures with large variation of local wavelengths and/or characteristic scales pose a challenge referred to as the mid-frequency problem. Standard numerical methods such as the finite element method (FEM) scale with the local wavelength and quickly become too large even for modern computer architectures. High frequency techniques, such as statistical energy analysis (SEA), often miss important information such as dominant resonance behaviour due to stiff or small scale parts of the structure.
Hybrid methods circumvent this problem by coupling FEM/BEM and SEA models in a given built-up structure.
In the approach adopted here, the whole system is split into a number of subsystems which are treated by either FEM or SEA depending on the local wavelength. Subsystems with relative long wavelengths are modeled using FEM.
Making a diffuse field assumption for the wave fields in the short wave length components, the
coupling between subsystems can be reduced to a weighted random field correlation function.
The approach presented results in an SEA-like set of linear equations which can be solved for the mean energies in the short wavelength subsystems.  }

\

{\noindent \it  \underline{keywords}: Hybrid methods, Statistical Energy Analysis, Wave chaos }

\


\section{Introduction}
\label{intro}

Determining the distribution of vibro-acoustic energy in complex built-up structures using finite element methods (FEM) is often a difficult task, in particular in the high frequency limit. The number of degrees of freedom (DoF) in FE models scale with the wavelength of the underlying wave field and computations quickly become prohibitive with increasing frequency. Moreover, ``numerically exact`` results obtained by an  FE approach for a specific structure may be of little practical value.  The vibro-acoustic response of ''identical'' structures assembled, for example, as part of a manufacturing process, is very sensitive to small changes in material parameters and/or variability of the shape of the structure. These differences may lead to large changes in the resonance spectrum and most of the detailed information of a full (and expensive) FEM calculation for an individual example has then at best statistical significance. In room acoustics, this computational ceiling may be as low as a few 100 Hz for moderate room sizes, it is roughly at 200 Hz for FE models describing the interior noise levels in cars \cite{Fahy, Kompella}. 

Statistical Energy Analysis (SEA) \cite{Lyon} is a popular tool to tackle the mid-to-high frequency regime. In an SEA framework, the whole system is viewed as a number of interacting subsystems giving rise to a set of linear equation for the mean wave energies stored in each subsystem. The SEA approach assumes that the wave field in each component is diffuse thus having universal statistical properties (that is, properties not depending on the shape, material parameters or even the wave equation characterizing a specific subsystem). SEA results may be interpreted as an average response of an ''ensemble'' of similar structures or of an individual system after averaging over frequency intervals large on the scale of the mean resonance density. SEA modelling is particularly useful if the wavelength is much smaller than characteristic scales of individual subsystems across the whole structure. 

In practical applications, this assumption often does not hold. Local wavelengths may vary considerably across a structure  and may at places become comparable to the size of a subsystem.  One may, for example, encounter stiff or narrow parts having only a few resonances in a given frequency band which are coupled to subsystems with a high modal density where SEA assumptions are valid. The long wavelength components will give rise  to characteristic resonance features which dominate the vibro-acoustic response - but can not be reproduced in an SEA approach. This leads to mid-frequency bands where the applicability of FEM and SEA do not overlap. To overcome such problems, a range of hybrid methods has been developed based on the concept of distinguishing between deterministic (stiff) and stochastic (short wavelengths) components. 

FEM is the method of choice for the deterministic subsystems while for the stochastic subsystems, one makes simplifying assumptions which reduce the computational cost and make it possible to find the response averaged over an  ensemble of statistical subsystems. One such hybrid method is the fuzzy structure theory \cite{Soize, Strasberg} which predicts the response of a master structure coupled to a large number of secondary structures. Assuming randomness in the secondary structures, the effects of the fuzzy substructures on the master structure can be described by  random impedance operators. This approach both reduces the number of DoF and gives an average response. Other hybrid methods include, (i) analytical impedance techniques \cite{Grice, Hong}, where floppy parts of the structure are modelled as receiver impedances; (ii) modal \cite{Gi} and SEA/modal \cite{Langley} approaches; and (iii) the ''Smooth Integral''/FEM hybrid method \cite{Pratellesi}, where probabilistic formulations are introduced as a modification of a boundary integral method.

A hybrid method based on wave concepts matching the solutions in the two kinds of subsystems without a priori assumptions about mode coupling was first developed by Shorter and Langley  \cite{Shorter, Cotoni}. The central idea of the method is a reciprocity result regarding the forces exerted on the boundaries of the deterministic subsystems \cite{Shorter1}. It is assumed that the wave field in stochastic subsystems can be decomposed into "direct" and "reverberant" field components and that the reverberant field component is described in terms of a random diffuse field. The reciprocity relationship couples the direct field radiation with diffuse reverberant loading at the interfaces between the deterministic and stochastic subsystems.

In this paper, we also divide the whole structure into stochastic and deterministic subsystems  and split the global wave field into a direct and a reverberant  component. We then introduce two important modifications from the Shorter and Langley approach  \cite{Shorter, Cotoni}. Firstly, we compute the {\em direct field explicitly and coherently across the whole structure} - details will be provided in Sec.\ \ref{sec:direct}. This yields  a direct coupling between deterministic subsystems which is  important in particular for medium to strong damping.  Secondly, we describe the response of  deterministic subsystems to external excitations in terms of Green functions. The coupling between the reverberant field in the deterministic and the stochastic subsystems is then facilitated by a {\em diffuse field-field correlation function} acting on the Green functions of the deterministic subsystems. The resulting expressions are formally equivalent to the reciprocity relation derived in \cite{Shorter1} using the force-force correlation function as has also been pointed out in \cite{Langley2}.  However, our approach simplifies the derivation considerably and offers an alternative viewpoint regarding possible future advancements of the method. The contributions of both wave fields will be finally coupled through energy balance equations. 
 
The article is organized as follows: the general set-up is introduced in Sec.\ \ref{sec:set-up}. In Sec.\ \ref{sec:direct}, we discuss a method to compute the direct field. In Sec.\ \ref{sec:stoch}, the average response of a deterministic subsystem to an energy influx from adjacent stochastic subsystems is described in terms of a random field correlation function. In Sec.\ \ref{sec:total},  we collect results from the previous sections and present the outline of the method. Numerical examples are given in Sec.\  \ref{sec:numeric}.

\section{The general set-up} 
\label{sec:set-up}
We will work with the Helmholtz equation in two spatial dimensions here, though the concepts are valid for general wave equations such as vectorial elasticity as well as for 3D problems. We will compute the solution to the wave equation for a set of coupled acoustic 2D cavities. To stay within an acoustics framework, we will use Neumann boundary conditions on the boundaries of the domains, although any other linear boundary condition could be used instead. 

\begin{figure} [t] 
\begin{center}
\includegraphics[width=0.47\textwidth, height=0.27\textwidth]{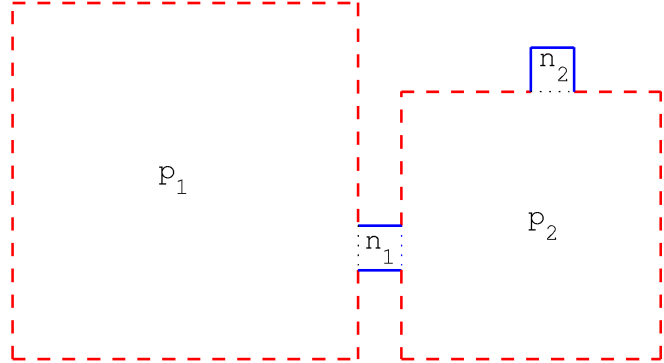}
\end{center}
\caption{(Colour online) The system is split into subsystems: $p_1 $ and $p_2 $  are stochastic subsystems, $n_1 $  and $n_2 $ are deterministic subsystems. Deterministic boundaries 
are shown as full lines and random boundaries as dashed lines. Dotted lines denote interfaces between subsystems. } \label{Fig1}
\end{figure}

The hybrid method is now set up as follows: the whole system is divided into {\em deterministic} and {\em stochastic} subsystems. Deterministic subsystems are those that, at a given frequency, are characterized by a wavelength comparable to the size of the subsystem. In actual applications, these may be bolts, frames, narrow joints or wave guides to name a few.  Stochastic subsystems are those for which the local wavelength is small relative to typical scales of the system. The wave field is generally characterized by rich interference patterns and its statistics is often well described by that of a  Gaussian random field or generalizations thereof \cite{Weaver1, TS07}. We assume that we know  the geometry of deterministic subsystems exactly and call the corresponding boundaries {\em deterministic boundaries}. For stochastic subsystems, we will not specify the actual form of the boundaries, only the area of the subsystem will enter the equations. We will often consider ensemble averages over stochastic subsystems by changing the boundary (keeping the area fixed) - consequently, we will call the 
boundaries of stochastic subsystems {\em random boundaries}, see also  \cite{Shorter1}. The ensemble average can be considered as averaging over a range of 'similar' systems, such as cars produced on an assembly line differing by small changes in fabrication. 

In the following, we assume that deterministic subsystems are connected to stochastic subsystems only. If two deterministic subsystems share a boundary, they are merged into one deterministic subsystem. The same holds for the stochastic subsystems. The whole system then consists of $N$ deterministic subsystems and $P$ stochastic subsystems, say. Fig.\ \ref{Fig1} shows a particularly simple example with $N = P = 2$; this system will be later considered for a numerical validation of the method. Our aim is to find the mean wave energies  in the stochastic  and deterministic subsystems, $\langle E_p \rangle$, $p=1,\ldots, P$ and
$\langle E_n \rangle$, $n=1,\ldots,N$, respectively.

We are looking here for approximate solutions of the stationary Helmholtz equation satisfying Neumann boundary conditions on all boundaries, that is,
\begin{equation}\label{helm} 
\sigma\bigtriangledown^2 \psi +  (\rho\omega^2 - i\eta \omega) \psi =f ,
\end{equation}
where $\psi$ denotes the acoustic pressure field over the whole system, $\eta$ is a damping coefficient, $\sigma$ is the stiffness, $\rho$ is the density and $f$ is a forcing term leading to the excitation of the system. In what fallows we set $\rho=1$. We will work with dimensionless units throughout; in the particular example, Fig.\ \ref{Fig1}, the width of the deterministic subsystem $n_1, n_2$ is scaled to one. 

We now decompose the global wave field into a {\em direct} and a {\em reverberant} field across the whole structure, that is, we write
\begin{equation} \label{decomp}\psi=\psi^{(d)}+\psi^{(r)}.\end{equation} 
Both these fields are defined {\em everywhere} in the structure, that is, both in the deterministic and stochastic
subsystems. The direct field $ \psi^{(d)}$ satisfies the wave equation (\ref{helm}) with Neumann boundary conditions on deterministic boundaries and {\em absorbing boundary conditions} on random boundaries. It thus corresponds to a wave field being reflected from and scattered by the boundaries of deterministic subsystems only, (full lines in Fig.\ \ref{Fig1}). Random boundaries, on the other hand, are ignored, that is, the system is considered open at these boundaries for the purpose of calculating $\psi^{(d)}$. For an example of a direct field, see Fig.\ \ref{Fig3}. 

The stochastic reverberant field  $\psi^{(r)}$ is the part of the total field
$\psi$ which originates from multiple reflections on both the random and deterministic boundaries. It can be considered as the wave field emerging after the first reflection of $\psi^{(d)}$ at the random 
boundaries. The boundary conditions for $\psi^{(r)}$ are  implicitly defined through the boundary conditions of the wave equation and value of the field $\psi^{(d)}$ on the random boundaries. As we will see later, there is actually no need to calculate $\psi^{(r)}$ explicitly in our hybrid method and results will be independent of the boundary conditions chosen on the random boundaries.
Ensemble averages will be performed by changing the boundaries of the stochastic systems leaving the area invariant; from the set-up, it is evident that these changes will only affect $\psi^{(r)}$, whereas the direct field $\psi^{(d)}$ remains independent of the shape of the random boundaries.  The reverberant field, on the other hand, is extremely sensitive to changes of these boundaries. Due to the size of the stochastic subsystem, small deformations of the boundaries will lead to large-scale changes in the interference patterns of $\psi^{(r)}$ originating from the multiple reflection processes. This opens up the possibility to describe the reverberant field in a statistical sense. The direct field is thus independent of the reverberant field $\psi^{(r)}$, but acts as a source term for the reverberant contribution $\psi^{(r)}$. \\


\section{Direct field}
\label{sec:direct}
We start by determining the direct field contribution $\psi^{(d)}$ of the global solution $\psi$ in Eq.\ (\ref{decomp}). The areas of the deterministic subsystems have been meshed to obtain FEM models for each subsystem. A part of the structure in  Fig.\ref{Fig1} together with the mesh is shown in Fig. \ref{Fig2} in more detail. Here, one can see the boundary of the structure, the mesh used in the FEM model of the deterministic subsystem and the interface which is introduced to separate the deterministic and the stochastic subsystem. We have a certain amount of freedom to chose this interface, that is, changing the form of the interface within the stochastic subsystem  will not change the overall result. Notice that in Fig. \ref{Fig2} the FEM mesh is somewhat expanded into the stochastic subsystem in contrast to the interface shown in Fig. \ref{Fig1}.

\subsection{Dynamic stiffness matrix for a single deterministic subsystem}
\label{sec:direct-one}
To start with, let us assume that there is only a {\em single}  deterministic subsystem, that is, we neglect any interaction between different deterministic subsystems. According to the definition given in the previous section, the direct field then satisfies Neumann boundary conditions on the deterministic boundaries and  transparent boundary conditions on the interfaces with the neighboring stochastic subsystems. Such a model accounts for the direct field radiation into stochastic parts, that is, outgoing waves emanating from interfaces of the deterministic subsystem will propagate outwards and will not return to the subsystem .
Describing the wave problem for  the deterministic subsystems in an FEM setting, the wave equation takes the form of a set of linear equations
\begin{equation} \label{1}
D^{(n)}\, {\bf q}^{(n)}= {\bf g}^{(n)},
\end{equation}
where $n$ is the deterministic subsystem under consideration. Here, ${\bf q}$ is the vector of nodal DoF, ${\bf g}$ is the vector of nodal forces, and  $D$ is dynamic stiffness matrix set up as usual in terms of stiffness, mass and damping matrices. 

There are various ways to implement transparent (absorbing) boundary conditions on the interfaces. One of the possible approaches to achieve absorbing boundary conditions is to construct an operator relating Dirichlet data to Neumann data on the interface. Such Dirichlet-to-Neumann map \cite{Givoli} can be cast in the form of what has been recently reported as hybrid FE - wave based technique in \cite{Genechten}. In particular, the application for a two dimensional unbounded problem is explained in Ref. \cite{Bergen}. In what follows we will use the method from Ref. \cite{Bergen} to construct the dynamic stiffness matrices of the deterministic subsystems. However, other methods such as {\em perfectly matched layer} (PML) techniques \cite{B87, BHPR07} could also be used.
 
In the following it will be advantageous to split the equation of motion for the deterministic subsystem into deterministic DoF ${\bf q}_d$ and interface DoF ${\bf q}_i$, that is,   
\begin{equation} \label{2}
\left( \begin{array}{ll}
D^{(n)}_d & D^{(n)}_{di} \\
 D^{(n)}_{id} & D^{(n)}_{i}
\end{array} \right)
\left( \begin{array}{l}
{\bf  q}^{(n)}_{d}\\
{\bf  q}^{(n)}_{i} 
\end{array} \right) =
\left( \begin{array}{l}
{\bf  g}^{(n)}_{d}\\
{\bf  g}^{(n)}_{i} 
\end{array} \right),
\end{equation}
where the subscripts $d$, $i$ denote DoF which lie inside deterministic subsystems and on interfaces, respectively. In particular, the sub-block $D_i$ accounts for absorbing boundary conditions for outgoing waves. In the setting considered here, the sources are in the deterministic subsystem and the DoF on the interface, ${\bf q}_i$, only contain outgoing wave components. We now move on to the case of several deterministic subsystems coupled through a direct field.

\subsection{The direct field propagator}
\begin{figure} [t] 
\begin{center}
\includegraphics[width=0.7\textwidth, height=0.42\textwidth]{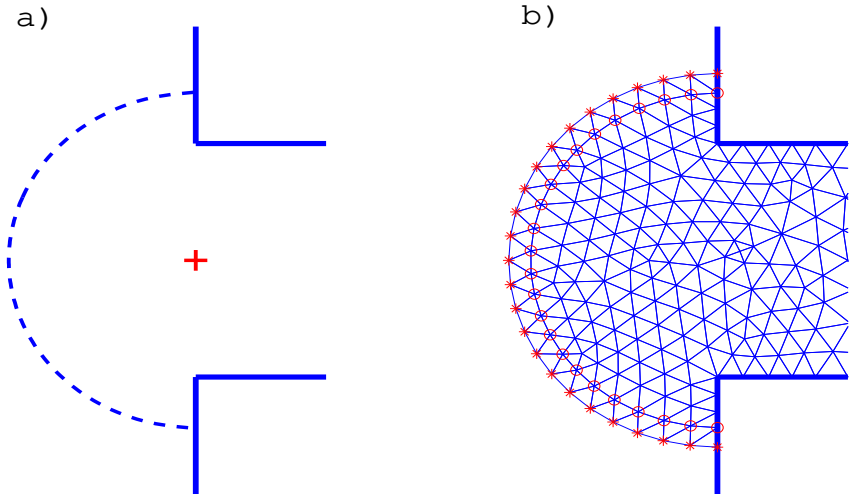}
\caption{(Colour online) a) An $(np)$ interface: solid lines are the boundary of the system, dashed lines, the interface between stochastic and deterministic subsystems. The interface is chosen in form of semicircle centred at $r_0$ (marked by a cross). b) The meshing for FEM computations: stars denote the degrees of freedom on the interface with field components ${\bf q}_{i}$, circles denote the degrees of freedom in the vicinity of the boundary inside the deterministic subsystem with field components $d$. 
} \label{Fig2}
\end{center}
\end{figure}

In a next step, we will describe the coupling between different deterministic subsystems in terms of free wave propagators mapping outgoing waves from one deterministic subsystem into incoming waves on another deterministic subsystem. In the following, we will use the notation $(np)$ for the interface between the $n_{th}$ deterministic subsystem and the $p_{th}$ stochastic subsystem assuming $n$ is adjacent to $p$. We denote by ${\bf q}_i^{(np)}$ the direct field components on the $(np)$ interface; the corresponding points are denoted with 'stars' in Fig.\ \ref{Fig2}. We will distinguish between incoming and outgoing components with respect to the deterministic subsystem on the $(np)$ interface and write
\begin{equation} \label{71}
{\bf q}^{(np)}_{i} = {\bf q}^{(np)}_{in}+{\bf q}^{(np)}_{out}.
\end{equation}
We will leave out the superscript $(d)$ (for direct field) in this section - all field quantities considered will be components of the direct field. 

We also restrict ourselves to circular interfaces. This choice is not a severe limitation of the method and allows for very general configurations due to the freedom we have to choose the interface between adjacent components. Let us now assume that we have found the wave solution in the deterministic subsystems. In particular, let us assume we know the outgoing wave component on the $(np)$ interface. 
(The actual coupling of FEM models corresponding to different deterministic subsystems will be described in the next section.) 
The direct field can now be extended into the stochastic subsystems. Due to the circular shape of the interface, we expand the outgoing field, ${\bf q}^{(np)}_{out}$, in terms of a Fourier basis, 
\begin{equation} \label{basis}
\phi_{m}^{(+)}(\theta) =\sin(2m\theta), \ \ \phi_{m}^{(-)}(\theta)=\cos(2m\theta),
\end{equation}
where $\theta$ is the angle parameterising the interface. That is, we write the outgoing wave field as 
\begin{equation}
\psi^{(np)}_{out}(\theta) =  \frac{1}{\pi} C_0 +  \frac{2}{\pi}\sum^{M}_{m=1}\left(C^{(+)}_{m}  \phi_{m}^{(+)}(\theta) +  C^{(-)}_{m}\phi_{m}^{(-)}(\theta)\right)
\end{equation}
with expansion coefficients 
\begin{equation} {\bf C}^{(np)} = {[C_0,   C^{(+)}_{1},  C^{(-)}_{1},  \dots,  C^{(+)}_{M}, C^{(-)}_{M}]}^{t}. \end{equation}
The wave field $\psi^{(np)}_{out}(\theta)$ corresponds to the wave vector $ {\bf q}^{(np)}_{out}$ in Eq.\ (\ref{71}) on the interface mesh points. 
The coefficients ${\bf C}^{(np)}$ can be obtained through numerical integration. The latter may be written in the form  
\begin{equation}
{\bf C}^{(np)}= V^{(np)} {\bf q}_{out}^{(np)}
\quad \mbox{with} \quad
 V_{m,l,\pm}^{(np)}=\phi_{m}^{(\pm)}(\theta_l)\Delta _l.
\end{equation}
The matrix $V$ performs the numerical integration over the interface to obtain the Fourier coefficients of the field ${\bf q}^{(np)}_{out} $ on the nodes of the FEM mesh. Here, the index $l$ labels the degrees of freedom on the interface and $\Delta_l$ is the summation weight of the quadrature. The truncation parameter  $M$ is determined by the wave length of the problem.
 
The direct field emanating from the $n_{th}$ deterministic subsystem into the $p_{th}$ stochastic subsystem through an $(np)$ interface can now be write explicitly as
\begin{equation} \label{bessels}
\psi^{(np)}({\bf r}) = \frac{C_0}{\pi H_{0}^{(1)}(kR) }H_{0}^{(1)}(kr)+  \frac{2}{\pi}\sum^{M}_{m=1}\frac{C^{(+)}_{m}  \phi_{m}^{(+)} +  C^{(-)}_{m}\phi_{m}^{(-)}}{H_{2m}^{(1)}(kR) }H_{2m}^{(1)}(kr),
\end{equation}
where $H^{(1)}_m$ are $m_{th}$ order Hankel functions outgoing with respect to the deterministic subsystem, ${\bf r} = (r, \theta)$ are the cylindrical coordinates centred at $r_{0}$ (cross in Fig.\ \ref{Fig1} a) and $R$ is the radius of the interface.  Note, that the wave number $k$ can be complex in case of non-zero damping. The solution written in the form of Eq.\ (\ref{bessels}) corresponds to radiation from the deterministic subsystem into the stochastic region; 
it will be used to evaluate solutions only on the interface and outside of the $n_{th}$ deterministic subsystem, that is, for $r\ge R$. Of course Eq. (\ref{bessels}) implies that the stochastic subsystem is homogeneous in a sense that the material parameters are constants.

The direct field emanating from the deterministic region $n$ into the stochastic subsystem $p$ adjacent to $n$ is fully described by the vector ${\bf C}$. 
The field at a point ${\bf r}$ inside the stochastic subsystem $p$ can now be obtained using Eq.\ (\ref{bessels}). That is, we obtain
\begin{equation} \label{L-op-def}
{\psi}^{(np)}({\bf r})= L^{(np)}({\bf r})\, {\bf q}_{out}^{(np)},
\end{equation}
where
\begin{equation}
\label{L-op}
L^{(np)}({\bf r}) ={\cal L}^{(np)}({\bf r}) V^{(np)}.
\end{equation}
and 
\begin{equation}
{\cal L}^{(np)}_{m, \pm}({\bf r}) =  \phi^{\pm}_m(\theta) \frac{H^{(1)}_{2m}(k r)}{H_{2m}^{(1)}(kR)}.
\end{equation}
In what follows, $L^{(np)}$ is termed the $(np)$ component of the {\em direct field propagator}.

To proceed we introduce a set of direct field propagators which are defined implicitly as  
\begin{equation} \label{70}
 {\bf q}^{(n'p)}_{in}=L^{(npn')} {\bf q}_{out}^{(np)}, 
\end{equation}
for $n\ne n'$. That is, $L^{(npn')}$ denotes the direct field propagator mapping outgoing waves on the $(np)$ interface onto incoming waves at the $(n'p)$ interface where the deterministic subsystems $n$ and  $n'$ are connected through subsystem $p$. Note, that $  {\bf q}^{(n'p)}_{in}$ corresponds to incoming wave contributions at the $(n'p)$ interface for $n\ne n'$. We define  $L^{(npn)} = 0$ which means that the direct field propagator produces only incoming waves with respect to the deterministic subsystems.  We can now define the total direct field propagators $L^{(p)}$ corresponding to the stochastic subsystem 
$p$ in the form  
\begin{equation} \label{700}
L^{(p)} =\left( \begin{array}{lll}
0 &\dots &L^{(Np1)}   \\
\vdots&  \ddots & \vdots \\
L^{(1pN)}& \cdots&  0\\
\end{array} \right).
\end{equation}
 Note that $L^{(p)}$ contains only those direct field propagators $L_{b}^{(npn')}$ with $n$ and $n'$ adjacent  to the stochastic subsystem $p$. A global direct field propagator is now defined as
\begin{equation} \label{701}
L =\left( \begin{array}{lll}
 L^{(1)}& &0 \\
       &\ddots  &\\
 0 & &     L^{(P)}\\ 
\end{array} \right). 
\end{equation}
Having constructed the global direct field propagator, Eq. (\ref{701}), one can related the incoming to the outgoing field components on all interfaces, that is,
\begin{equation}\label{Ldir}
 {\bf q}_{in} = L \, {\bf q}_{out}. 
\end{equation}


\subsection{The global equation of motion for the direct field}
We saw in the previous subsection that the direct field in the stochastic subsystems is determined by the solution on the interfaces alone, as we do not account for any reflections from random boundaries. Our treatment thus automatically fulfills the free radiation (or absorbing) boundary conditions on the random boundaries. 
We proceed by constructing  the direct field solutions in the deterministic subsystems and on the interfaces on an FE mesh making use of the direct field propagator. 

Using Eq.\ (\ref {Ldir}), one can couple Equations of motion (\ref {2}) for individual deterministic subsystems to obtain
\begin{equation} \label{dir_field}
\left( \begin{array}{ll}
D_d &  D_{di}(I + L) \\
D_{id} & {D}_{i}
\end{array} \right)
\left( \begin{array}{l}
{\bf  q}_{d}\\
{\bf  q}_{out}
\end{array} \right) =
\left( \begin{array}{l}
{\bf  g}_{d}\\
{\bf  g}_{i}
\end{array} \right), 
\end{equation}
where $I$ is the identity operator on the boundary DoF. The matrix $D_d$ is a block diagonal matrix set up by the sub-matrices $D^{(n)}_d$ defined in Eq.\ (\ref{2}) and similarly for $D_{di}, D_{id}, D_i$; that is, all matrices and vectors in (\ref{dir_field}) are defined globally across all deterministic subsystems and their interfaces. Note, that $D_i$ accounts for absorbing boundary conditions for the outgoing fields at interfaces.
Incoming waves responsible for the coupling between wave fields in different  deterministic substructures are taken into account through the direct field propagator $L$ via Eq.\ (\ref{Ldir}).

\begin{figure} [t] 
\begin{center}
\includegraphics[width=0.5\textwidth, height=0.35\textwidth]{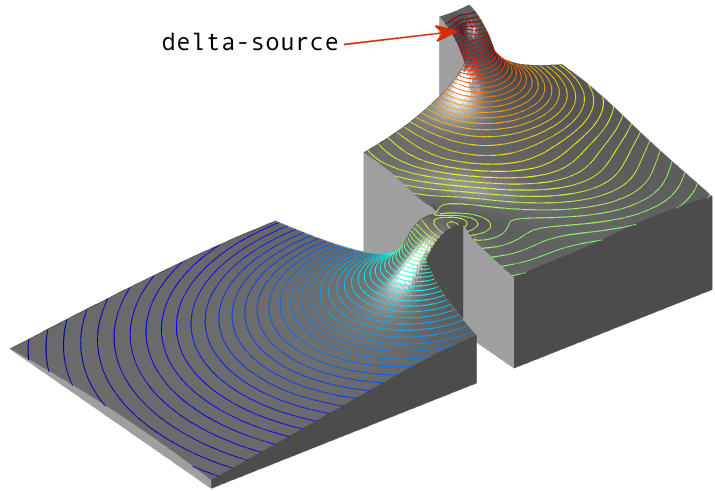}
\caption{(Colour online) The logarithmic plot of a solution for the energy in the direct field.} \label{Fig3}
\end{center}
\end{figure}

Equation \ (\ref{dir_field}) is now solved for ${\bf q}_d$ and ${\bf q}_{out}$. The incoming component ${\bf q}_{in}$ can then be obtained from  Eq.\ (\ref{Ldir}). The solution of Eq.\ (\ref{dir_field}) gives the wave field produced inside the deterministic subsystems and on the interfaces including coupling of different deterministic subsystems via radiation flowing through stochastic subsystems.  Damping in stochastic subsystems is taken into account by choosing complex wave numbers in the direct field propagator $L$. Note that the direct field is coherent across the whole structure and that the deterministic subsystems exchange energy through the direct field. If we have only one deterministic subsystem as considered in Sec.\ \ref{sec:direct-one}, the method introduced is an implementation of absorbing boundary conditions on the circular interfaces as for example considered in \cite{Bergen, Li}. The direct field can be found everywhere in the stochastic region with the help of the direct field operator $L$ defined in Eqs.\ (\ref{L-op-def}), (\ref{700}), (\ref{701}). A solution for the direct field in the structure shown in Fig. \ref{Fig1} is presented in Fig. \ref{Fig3}. It can clearly be seen that the direct field  propagates freely through the random boundaries.

We have so far assumed that the structure is excited in a deterministic subsystem. If the forcing is performed inside a stochastic subsystem $p$, we have to add to the right hand side of Eq.\ (\ref{dir_field}) a force acting on the $(np)$ interfaces by the incoming wave field emanating from the source. The latter can be calculated using the free field Green function inside $p$.   

In order to compute the energy flux transported via the direct field from deterministic to stochastic subsystems one needs the direct field components  ${\bf q}_{d}$ and the total field on the boundary ${\bf q}_{i} $. They can be obtained from the solution of Eq.\ (\ref{dir_field}) together with Eq.\ (\ref{Ldir}). The direct field energy input into the $p_{th}$ stochastic subsystem per period $T= 2\pi/\omega$ is then given as
\begin{equation} \label{flux_1}
 P_{p}^{(d)}=\frac{\omega}{2}\sum_{n=1}^{N}  \Im{\{} ({\bf q}^{(np)}_{i})^{\dagger} \Re\left\{ D_{id}^{(np)} \right\} {\bf q}^{(np)}_{d} {\}}, 
\end{equation}
where $D_{id}^{(np)}$  is  sub-matrix of $D_{id}$ which account for interactions at the $(np)$ interface. In fact, to use Eq. (\ref{flux_1}) one only needs to evaluate ${\bf q}^{(np)}_{d}$ at mesh points inside $n$  next to the interface (marked by circles in Fig.\ \ref{Fig2}).

An extension of the method to 3D can be performed along the same line. In this case, the interfaces have the form of spherical segments. The segments of  a sphere defining the interface can be of fairly arbitrary shape. Finding a 2D, orthogonal basis on these interfaces in analogy to the Fourier basis in Eq. (\ref{basis}) will lead to an additional numerical overhead. However, the general principle remains the same.

\section{The stochastic reverberant field}
\label{sec:stoch}

\subsection{FEM equations}

To account for the response of the stochastic subsystems, we need to estimate the stochastic reverberant field $\psi^{(r)}$ in addition to the solution of Eq.\ (\ref{dir_field}). The stochastic field is created in interference of waves reflected from the random boundaries. In what follows, we will assume that  $\psi^{(r)}$ is  a random diffuse field, see \cite{Weaver1} and references therein. We want to avoid an exact computation of $\psi^{(r)}$ either because we do not have enough information about the random boundaries or because a numerically exact treatment of the stochastic systems would need exorbitant large FEM models. In our hybrid approach, we will thus only take into account the response of a deterministic subsystem to the reverberant field exited in the neighboring stochastic subsystems. We will furthermore assume that the reverberant fields in different stochastic subsystems are uncorrelated and that the reverberant field inside the $n_{th}$ deterministic subsystems is an incoherent superposition of the partial fields excited through the different  interfaces of $n$. It is thus sufficient to compute the response of the $n_{th}$ deterministic subsystem to a diffuse excitation at every $(np)$ interface separately for all $p$ adjacent to $n$.  

The stochastic  field in subsystem $p$ leads to stochastic field components on the $(np)$ interfaces which can be modelled in terms of stochastic reverberant forces. These forces in turn excite vibrations in the deterministic subsystems $n$ and lead to transport of vibro-acoustic energy through subsystem $n$ to other stochastic subsystems $p'$ adjacent to $n$. We will denote the reverberant field in the $n_{th}$ deterministic subsystem as ${[{\bf q}_d^{(r)}, {\bf q}_i^{(r)}]}^{t}$ and the discussion will be restricted to the local field in $n$ from now on. We will also omit the superscript $(r)$ for reverberant field in this section - {\em all field quantities considered will be reverberant}. 

We start by computing the reverberant field in a deterministic subsystem $n$ excited on the $(np)$ interface by a stochastic field in subsystem $p$. We again split the solution on the interfaces (here of subsystem $n$ only) as
$${\bf q}^{(n)}_i ={\bf q}^{(n)}_{out}+ {\bf q}^{(n)}_{in} $$
and write the FEM equations in the form
\begin{equation} \label{rev_field}
\left( \begin{array}{ll}
D^{(n)}_d & D^{(n)}_{di}  \\
D^{(n)}_{id} & {D}^{(n)}_{i}
\end{array} \right)
\left( \begin{array}{l}
{\bf  q}_{d}\\
{\bf  q}_{out}
\end{array} \right) =
\left( \begin{array}{l}
 {\bf f}^{(np)}\\
{\bf  0}
\end{array} \right)
\end{equation}
with the same dynamic stiffness matrices as defined in Eq.\ (\ref{2}). It implements absorbing boundary conditions at the interfaces, that is, the interfaces act as openings through which outgoing waves leave the system without being reflected.
The right hand side of Eq.\ (\ref{rev_field}) contains the stochastic reverberant force
\begin{equation} \label{222}
{\bf f}^{(np)}=-D_{di}^{(np)}{\bf  q}^{(n)}_{in}
\end{equation}
acting on the deterministic DoF next to the $(np)$ interface (marked by circles in Fig.\ \ref{Fig2}). Note that the only non-zero entries of the force vector ${\bf f}^{(np)}$ are those along the $(np)$ interface. We of course do not know ${\bf q}_{in}$, so this variable needs to be eliminated eventually.  

The solution of Eq.\ (\ref{rev_field}) for a given force term can be found with the help of the Green function $G^{(np)}$ acting on the DoF inside subsystem $n$ with source points on the $(np)$ boundary. It is the solution of the equation
\begin{equation}\label{Green}
\left( \begin{array}{ll}
D^{(n)}_d & D^{(n)}_{di}  \\
D^{(n)}_{id} &{D}^{(n)}_{i}
\end{array} \right)
\left( \begin{array}{l}
 G^{(np)}_{d} \\
 G^{(np)}_{i} 
\end{array} \right) =
\left( \begin{array}{l}
I^{(np)}_{\circ} \\
0
\end{array} \right),
\end{equation}
where $I^{(np)}_{\circ}$ is the identity matrix on the DoF next to the $(np)$ interface marked by circles in Fig.\ \ref{Fig2}. 

The Green function in Eq.\ (\ref{Green}) accounts for the response of the subsystem $n$ to a force acting on the DoF next to the $(np)$ interface.  We define the part of the Green function $G^{(np)}$ which gives the wave field on an $(np')$ interface due to an excitation on the $(np)$ interface as $G^{(pnp')}$. The sub-matrices $G^{(pnp')}$ can now be calculated by FEM. 

\subsection{Flux formulas}

The energy flux from the $(np)$ to the $(np')$ interface produced by the force ${\bf f}^{(np)}$ in the $n_{th}$ deterministic subsystem can be written in a form similar to Eq.\ (\ref{flux_1}), that is,
\begin{equation} \label{flux_2}
 P^{(n)}_{p\rightarrow p'}=\frac{\omega}{2} \Im{\{}  ({\bf f}^{(np)})^\dagger  
{(G_{i}^{(pnp')})}^{\dagger} \Re\left\{D_{id}^{(np')} \right\} 
G_{d}^{(pnp')} {\bf f}^{(np)}  {\}}.
\end{equation}
One can also find the total flux of energy flowing from the 
$p_{th}$ stochastic subsystem into the $n_{th}$ deterministic subsystem, that is,
\begin{equation} \label{flux_3}
 P_{p\rightarrow n}=\frac{\omega}{2} \Im{\{}  {\{{\bf q}^{in} + G_{i}^{(pnp)} {\bf f}^{(np)}}\}^{\dagger}  \Re\left\{D_{id}^{(np)} \right\}
G_{d}^{(pnp)} {\bf f}^{(np)}  {\}}.
\end{equation}

Note that
\[ \sum_{p=1}^P P_{p\rightarrow n} \ne \sum_{p,p'=1}^P P^{(n)}_{p\rightarrow p'}\]
in general; equality holds only if there is no energy loss due to, for example, damping in subsystem $n$. 

So far everything is exact; however, the stochastic force given in Eq.\ (\ref{222}) is not known as in depends on ${\bf q}_{in}$. We proceed now by taking averages over the power inputs in terms of the reverberant field components. That is, we consider averages $\langle P \rangle$ by determining fluxes  over an ensemble of stochastic subsystems where we make small changes to the random boundaries. Using Eqs.\ (\ref{222}) and (\ref{flux_2}), one can write  
\begin{equation} \label{77}
 \langle P^{(n)}_{p\rightarrow p'} \rangle =
\frac{\omega}{2}  \Im{\{} \sum_{j,k} T_{j,k}^{(p\rightarrow p',n)} \langle {(q^{j}_{in})}^{*}q^{k}_{in} \rangle {\}},
\end{equation}
where
\begin{equation} \label{88}
T^{(p\rightarrow p',n)}= D_{di}^{\dagger}  {\{G^{(pnp')}_i\}}^{\dagger}   
 \Re\left\{D_{id}^{(np')} \right\}G^{(pnp')}_d D_{di},
\end{equation}
and indexes $j, k$ run over DoFs on the corresponding interface.
The average energy influx from the $p_{th}$ to the $n_{th}$ subsystem can be written in the form 
\begin{equation} \label{777}
{ \langle P}_{p \rightarrow n} \rangle =
\frac{\omega}{2} \Im{\{} \sum_{j,k} T^{p \rightarrow n}_{j,k} \langle {(q^{j}_{in})}^{*}q^{k}_{in} \rangle {\}},
\end{equation}
where, according to Eq.\ (\ref{flux_3}), 
\begin{equation} \label{888}
T^{p \rightarrow n}= \left(I+ D_{di}^{\dagger}  {\{G^{(pnp)}_i\}}^{\dagger} \right) \Re\left\{D_{id}^{(np)} \right\} G^{(pnp)}_d  D_{di}.
\end{equation}
Here, the identity matrix accounts for the incoming wave energy, whereas outgoing flux on the $(np)$ interface due to reflections inside the $n_{th}$ subsystems is described by the Green function  $G^{(pnp)}_i$.


\subsection{The field-field correlation function}
The field variables ${\bf q}_{in}$ enter now only in form of the correlation functions 
$\langle {(q^{j}_{in})}^{*}q^{k}_{in} \rangle$. The latter can be express
in the following form
\begin{equation} \label{9}
\langle {(q^{j}_{in})}^{*}q^{k}_{in} \rangle = C_{p}^2 F({\bf r}_{j},{\bf r}_{k}),
\end{equation}
where  $F({\bf r}_{j},{\bf r}_{k})$ is the area (volume) normalized field amplitude correlation function, and $C_{p}^2$ is the mean squared amplitude of the incoming component of the random wavefield in the $p_{th}$ subsystem. Taking into account that incoming and outgoing components (with respect to any direction) of the diffuse field store the same amount of energy, one can find the time and ensemble averaged kinetic energy in the $p_{th}$ subsystem as 

\begin{equation}
\langle E_p^{(kin)} \rangle= \frac{\rho_{p} \omega^2 C_{p}^2 S_p}{2},
\end{equation}
where $\rho_{p}$ is the material density and $S_p$ is the area (volume) of the subsystem. Finally, using the virial theorem, one can express $C_{p}^2$ through the total energy

\begin{equation} \label{101}
C_{p}^2=\frac{ \langle E_p \rangle}{\rho_{p} \omega^2 S_p}. 
\end{equation}
The problem is thus reduced to finding the correlation function $F({\bf r}_{j},{\bf r}_{k})$.  The issue of random field correlation function was discussed in \cite{Weaver1} in the context of random diffuse field theory or in \cite{Urbina} in the context of quantum and wave chaos. The central result is that the correlation function for a diffuse field is (again after averaging over an ensemble of similar systems) equal to the imaginary part of the Green function of this system \cite{Weaver2, Gouedard}. In particular, the area-normalized correlation function is given in the following form
\begin{equation}\label{105}
 F({\bf r}_j, {\bf r}_k)
= \frac{ S_p }{  \Im \{ \mathrm{tr} \left( G({\bf r}_{ j},{\bf r}_{k}) \right) \} } \Im \{  G({\bf r}_{ j},{\bf r}_{k}) \},
\end{equation}
where $G({\bf r}_{ j},{\bf r}_{k})$ is he Green function connecting points ${\bf r}_{ j},{\bf r}_{k}$ and ${\rm tr} \,G$ denotes the trace.
Taking into account that the trace of the Green function is related to the modal density, one can write
\begin{equation} \label{102}
 n(\omega)= -\frac{2 \omega \rho_p }{\pi} \Im \{ \mathrm{tr} \left( G({\bf r}_{ j},{\bf r}_{k}) \right) \}. 
\end{equation}
The modal density of the subsystem is to leading order given by  the area (volume) of the subsystem; in the two dimensional case the expression for the modal density is thus of the form 
\begin{equation} \label{Weyl}
 n(\omega)=\frac{S_p \omega}{2\pi c^2},
\end{equation}
where {$c$ is the speed of sound.} Corrections to the leading order behavior for the modal density due to, for example, the boundary of the subsystem can be considered in terms of a Weyl expansion, see \cite{TS07, Urbina}. Finally, using  Eqs.\ (\ref{105}), (\ref{102}), one finds
\begin{equation} \label{99}
\langle {(q^{j}_{in})}^{*}q^{k}_{in} \rangle = -\frac{ 2 \langle E_p \rangle}{  \pi \omega n(\omega) } \Im \{  G({\bf r}_{ j},{\bf r}_{k}) \}.
\end{equation}

It may seem that the result obtained does not simplify the problem, because we do not know the Green function for the whole system. However, we can approximate the global solution locally using the free-space Green function. In fact, this issue was thoroughly investigated in wave chaos theory, see \cite{TS07, Kuhl} for a reviews. It turns out that the imaginary part of the free-space Green function describes the correlations in Gaussian random fields correctly. The free-space Green function can indeed be viewed as the zeroth-order approximation to the true correlation function \cite{Shubert}. For the 2D Helmholtz equation considered here, the free Green function is given as 
\begin{equation}
 G_0({\bf|r|})= \frac{1}{4} Y_{0}(\sqrt{z}{\bf|r|}) - \frac{i}{4}  J_{0}(\sqrt{z}{\bf|r|}) \},
\end{equation}
where the complex part of  $z=-i\eta\omega+\omega^2$  takes into account absorption. In the case $ \eta=0$ and using the free Green function, yields the well known result for Gaussian random waves \cite{Berry}
$$F(|{\bf r}|)=J_{0}(\omega|{\bf r}|) .$$

The average flux in Eqs.\ (\ref{77}) and (\ref{777}) is found by using a uniform diffuse field assumption with uniform reverberant field amplitudes $C_p$ across the boundary. This is a basic assumption of SEA and the coefficients $T^{(p\rightarrow p',n)}$ can be related to SEA coupling constants. Moreover, we assumed that the diffuse field at different interfaces of a given deterministic subsystem are not correlated. The random boundaries in each subsystem are indeed assumed to vary independently; if furthermore the lengths of the random boundaries of the stochastic subsystems exceed the lengths of the interfaces, the reverberant field in each subsystem is most strongly influenced  by its boundaries. Thus, one can neglect coherent power transfer between two stochastic subsystems {\cite{Mace}} and the average net flux carried by the diffuse field through any section of a deterministic subsystem is the sum of contributions given by Eqs.\ (\ref{77}),  (\ref{777}). 

\section{Energy balance equations}
\label{sec:total}

We have now all the ingredients to determine the total energy influx into the $p_{th}$ stochastic subsystem through its interfaces, that is, 
\begin{equation} \label{111}
 \langle P_{p}^{in} \rangle=  P_{p}^{(d)} +
  \frac{1}{2 \omega} \sum_{p'=1}^{P}  \frac{A_{p'}^{p} \langle E_{p'}^{(r)} \rangle }{\rho_{p'} S_{p'}},
\end{equation}
where the direct field contribution $P_p^{(d)}$ is given in (\ref{flux_1})  and 
\begin{equation} \label{122}
 A_{p'}^{p}=\sum_{n=1}^{N}  \Im{\{} \sum_{j,k} T_{j,k}^{(p'\rightarrow p,n)} F({\bf r}_{j},{\bf r}_{k}) {\}};
\end{equation}
note, that $A_{p'}^{p}$ is zero if the $p'_{th}$ subsystem is not directly connected to 
the $p_{th}$ subsystem via a deterministic subsystem.

The mean energy flowing out of the $p_{th}$ subsystem either due to damping and or due to 
energy transfer into neighboring deterministic subsystems can be written as
\begin{equation} \label{13}
 \langle P_{p}^{out} \rangle= \frac{\omega \eta_p}{2} \int\limits_{S_p}^{} 
{({\psi}_{p}^{(d)})}^{*}{\psi}_{p}^{(d)} dS_p + \frac{ \eta_{p} \langle E_{p}^{(r)} \rangle} { \rho_p } +
  \frac{1}{2 \omega}  \frac{B_{p} \langle E^{(r)}_{p} \rangle }{\rho_{p} S_{p}},
\end{equation}
where 
\begin{equation} \label{123}
 B_{p}=\sum_{n=1}^{N}  \Im{\{} \sum_{j,k} T_{j,k}^{(p\rightarrow n)} F({\bf r}_{j},{\bf r}_{k}) {\}}
\end{equation}
and $\eta_{p}$ is the damping factor. We assume here that $\eta_{p}$ is constant across each stochastic subsystem. The first term in Eq.\ (\ref{13}) is the energy lost due to damping in the direct field $\psi^{(d)}$ which can be found with the help of appropriate direct field propagators $L$. The second term is the energy lost in subsystem $p$ due to damping and the last term gives the energy transferred into adjacent deterministic subsystems.

We finally obtain a set of SEA-like equations using the energy balance relations 
$\langle P_{p}^{in} \rangle=\langle P_{p}^{out} \rangle$. That makes it possible to reduce the problem 
to a set of linear equations 
\begin{equation} \label{14}
\left\{
\begin{array}{l}
  \langle P_{1}^{in} \rangle-\langle P_{1}^{out} \rangle=0       \\
  \langle P_{2}^{in}\rangle- \langle P_{2}^{out}\rangle=0       \\
      \ \ \dots                  \\
  \langle  P_{P}^{in}\rangle-\langle P_{P}^{out}\rangle=0
\end{array}
\right.  
\end{equation}
which can be solved for the unknowns $\langle E_{1}^{(r)}\rangle,  \langle E_{2}^{(r)}\rangle, 
\dots, \langle E_{P}^{(r)}\rangle$. The energies stored in the direct field  $\langle E_{1}^{(d)}\rangle,  
\dots, \langle E_{P}^{(d)}\rangle$ can again be found with the help of the direct field propagator $L$.

We now rewrite the set of equations in matrix form, that is,
\begin{equation} \label{matrix}
\left( \begin{array}{ccc}
 M_{11}-\frac{\eta_1}{\rho_1} &\cdots & M_{1P}   \\
\vdots&  \ddots & \vdots \\
M_{1P} & \cdots&  M_{PP}-\frac{\eta_P}{\rho_P}   \\
\end{array} \right)
\left( \begin{array}{c}
\langle E_1 \rangle   \\
\vdots \\
 \langle E_P \rangle   \\
\end{array} \right) 
=
\left( \begin{array}{c}
Q_1  \\
\vdots \\
 Q_P \\
\end{array} \right), 
\end{equation}
where the diagonal terms $M_{pp}$ are given as
\begin{equation}
 M_{pp} = \frac{1}{2 \omega}  \frac{B_{p}}{ \rho_{p} S_{p}},
\end{equation}
the off-diagonal terms $M_{pp'}$ have the form
\begin{equation}
 M_{pp'} = -\frac{1}{2 \omega}  \frac{A_{p'}^{p}}{ \rho_{p'} S_{p'}},
\end{equation}
and the source terms are 
\begin{equation}
 Q_p = P^{(d)}_p-\frac{\omega \eta_p}{2} \int\limits_{S_p}^{} 
{({\psi}_{p}^{(d)})}^{*}{\psi}_{p}^{(d)} dS_p.
\end{equation}
We can now directly read off coupling loss factors and similar SEA parameters from Eq.\ (\ref{matrix}). Note that in contrast to traditional SEA, we can now incorporate losses due to damping within the deterministic subsystems and the summation over the row vectors in (\ref{matrix}) may not add up to zero.

One can also find the reverberant energies in the $n_{th}$ deterministic subsystem as an incoherent sum of reverberant contribution from all neighboring subsystems
\begin{equation}
\langle E^{(r)}_{n} \rangle  = \sum_{p} \langle E^{(r)}_{(np)}  \rangle,
\end{equation}
where
\begin{equation}
 \langle E^{(r)}_{(np)} \rangle = \frac{-\langle E_{p}^{(r)} \rangle}{2 \pi \omega n_p(\omega)} \sum_{j,k} W^{(np)}_{j,k} \Im \left\{ G_0({\bf r}_j, {\bf r}_k)\right \},
\end{equation}
and
\begin{equation}
 W^{(np)}={D_{di}}^{\dagger} {(G_{i}^{(np)})}^{\dagger} \Re \left \{ D^{(n)} \right \}
 G_{d}^{(np)} D_{di}.
\end{equation}

Our approach is thus a variant of SEA with respect to treating the stochastic components. The coupling loss factors contain wave information from the deterministic subsystems  - they keep track of resonances excited in the deterministic subsystems, but  coupled to the stochastic components. In particular, our hybrid method provides 

(i) direct coupling between deterministic subsystems via the direct field which is here calculated explicitly and coherently across the whole structure - a contribution not considered so far in hybrid approaches.

(ii) an exact numerical treatment of the deterministic subsystems giving rise to coupling between stochastic subsystems. The method thus yields frequency dependent  SEA coupling parameters. 

(iii) a statistical treatment of the stochastic subsystems.  Using the correlation function
$$ F({\bf r}_{j},{\bf r}_{k}) = \langle \psi^{*}({\bf r}_j) \psi({\bf r}_k) \rangle $$
is equivalent to the assumption that the underlying reverberant field is diffuse. Our approach is equivalent to using the direct field reciprocity relation derived in \cite{Shorter1}; it gives, however, a more straightforward access to the problem.

\section{Numerical validation}
\label{sec:numeric}

\begin{figure} [h] 
\begin{center}
\includegraphics[width=0.45\textwidth, height=0.45\textwidth]{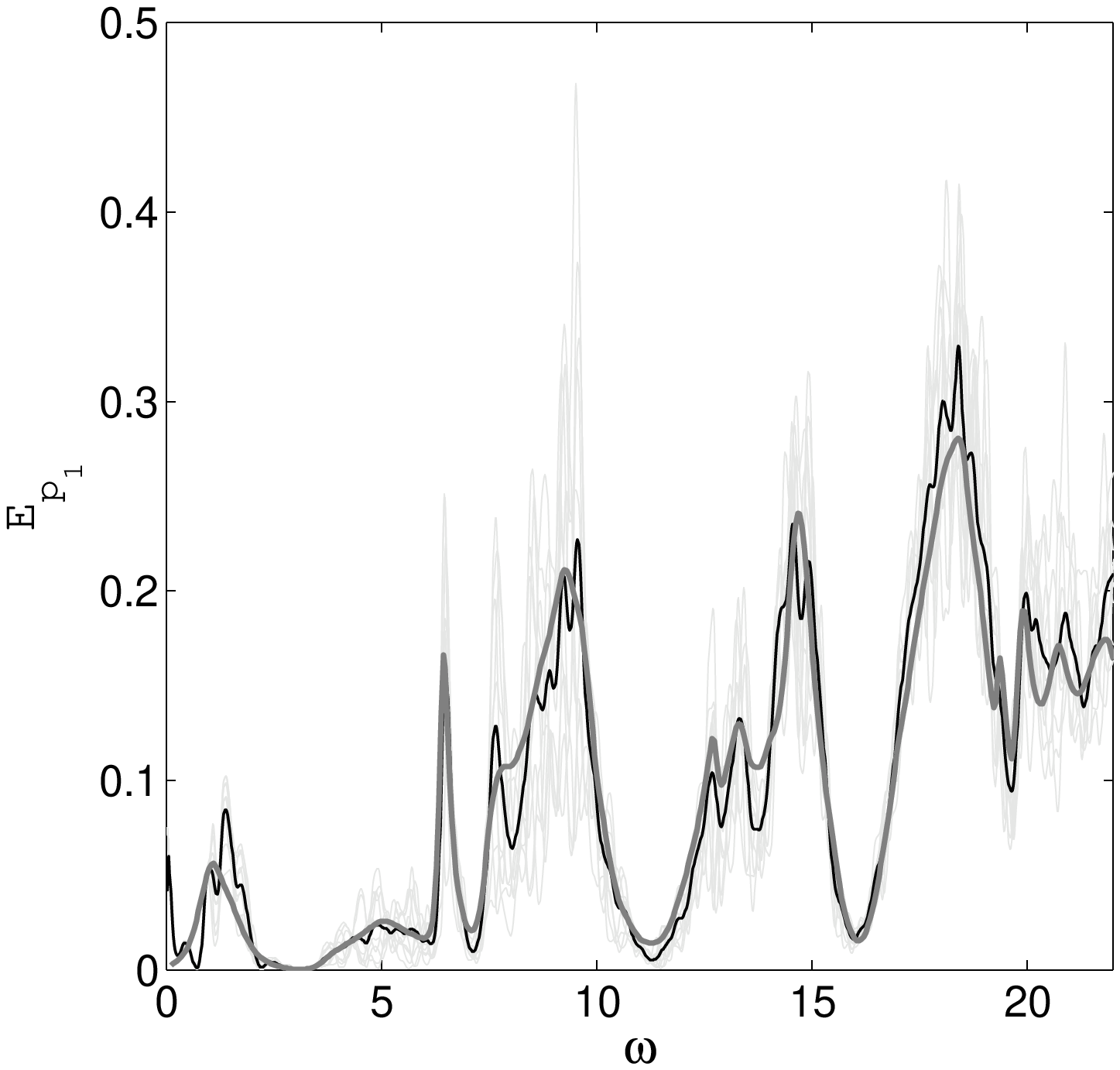}
\includegraphics[width=0.45\textwidth, height=0.45\textwidth]{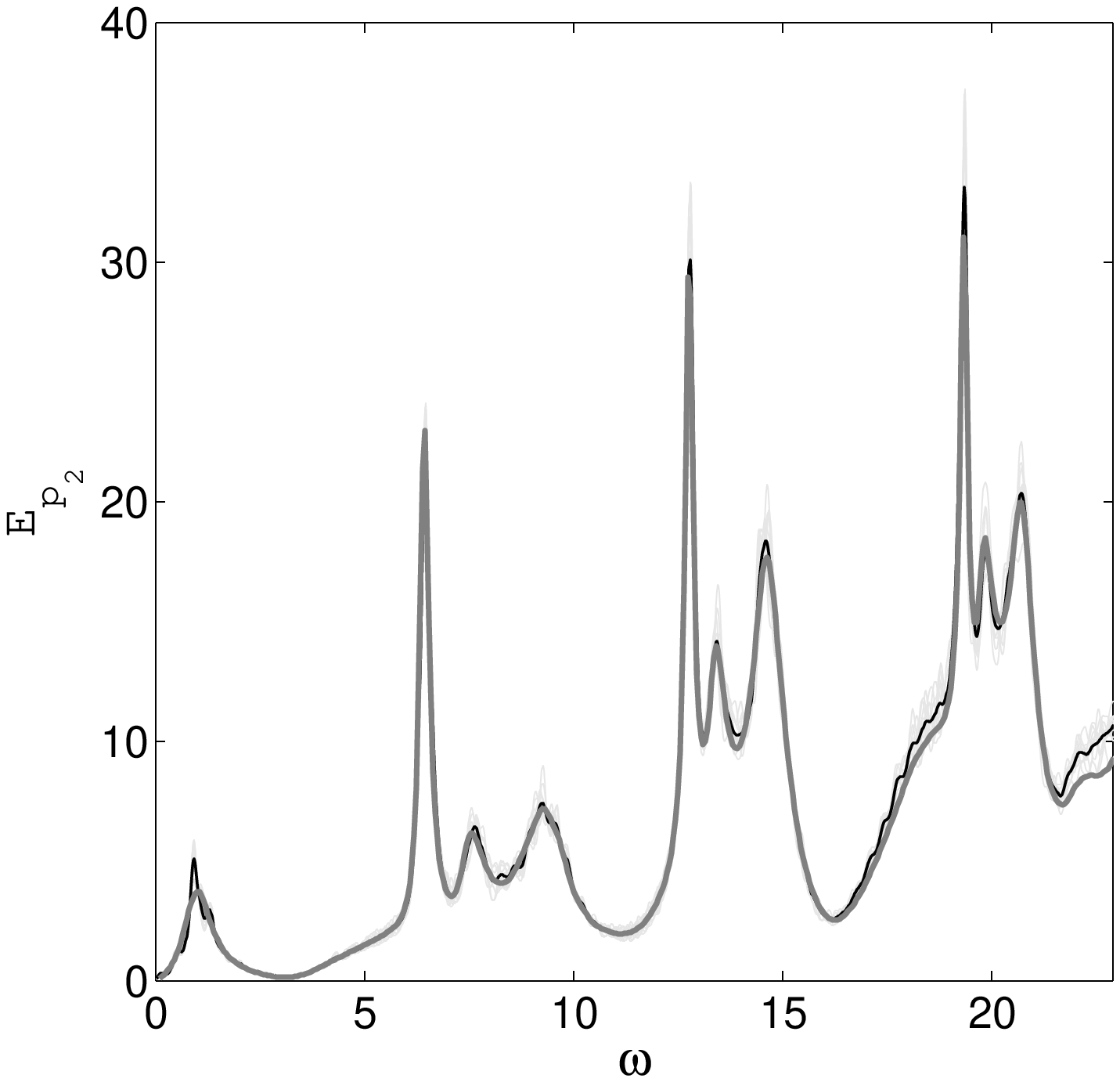}

\caption{The energies in the left ($p_1$) and right ($p_2$) region found with FEM using Monte Carlo sampling over randomized boundaries compared to the results obtained with the hybrid method. Light grey: each Monte Carlo realization; black: ensemble average; bold grey: hybrid method; $\eta=0.20$.} \label{Fig4}
\end{center}
\end{figure}
The method has been numerically validated for the structure shown in Fig.\ \ref{Fig1} with  a $\delta$-function force in one of the deterministic subsystem as shown in  Fig.\ \ref{Fig3}.  We solve the wave problem using both the hybrid approach and by directly applying the FEM. To introduce randomness into the latter, we slightly distort the shape of the  boundaries in the stochastic subsystems keeping the areas constant and find the FEM solution for every realization. Depending on the damping parameter, there are two main scenarios. In the {\em high damping case}, the energy flux into the stochastic subsystems is damped before it reaches the random boundaries and reflections into the  reverberant field are suppressed. A large part  of the energy is thus stored in the direct field. In the  {\em low damping case}, 
energy loss in the direct field is insignificant compared to the net flux into the stochastic components and most of the energy is stored in the stochastic reverberant field. In this case, almost all waves reach the random boundaries and are reflected into the reverberant field where they accumulate. Increasing the damping parameter $\eta$, one can cross over from one to the other regime. 

\begin{figure} [t]
\begin{center}
\includegraphics[width=0.45\textwidth, height=0.45\textwidth]{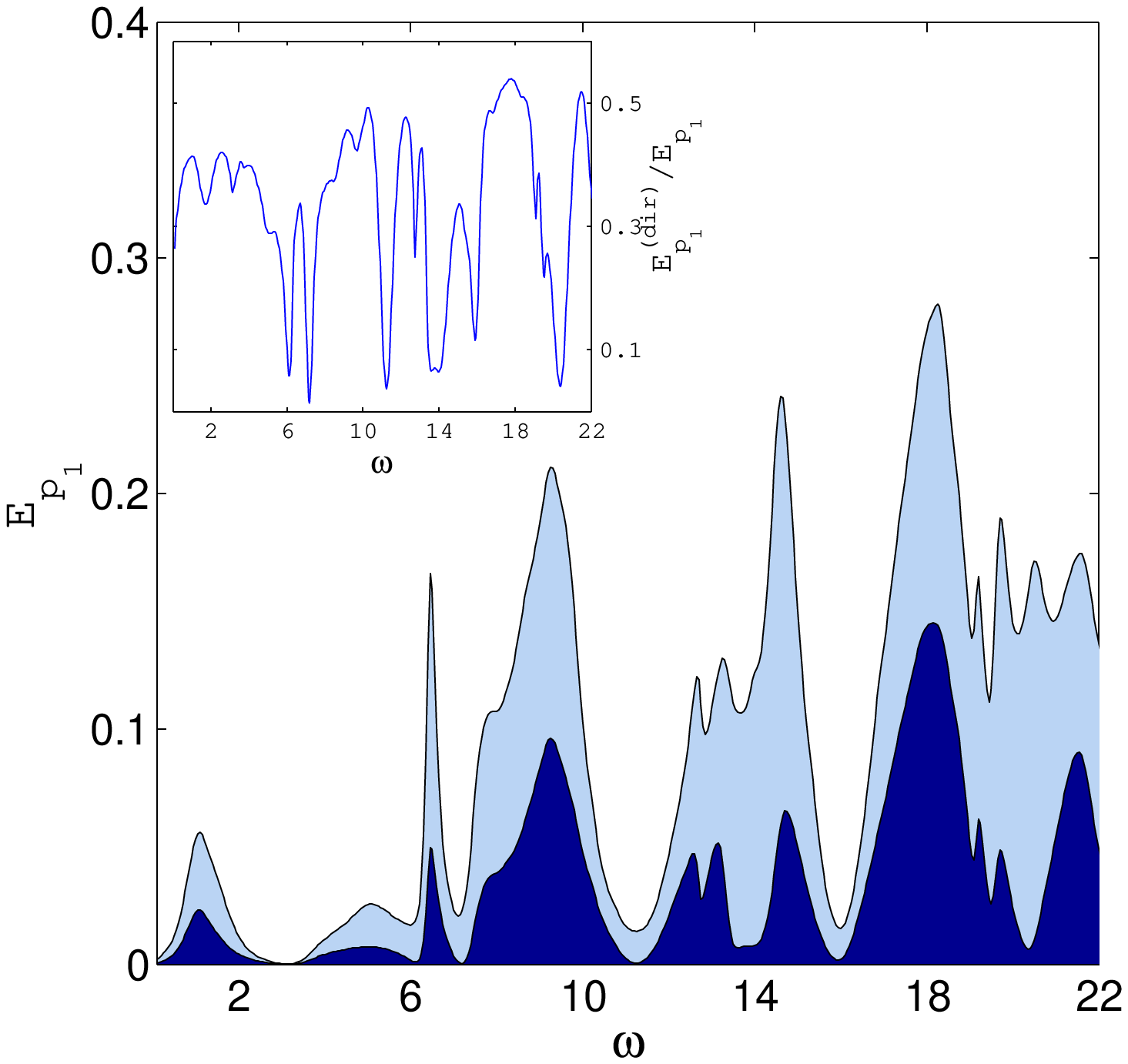}
\includegraphics[width=0.45\textwidth, height=0.45\textwidth]{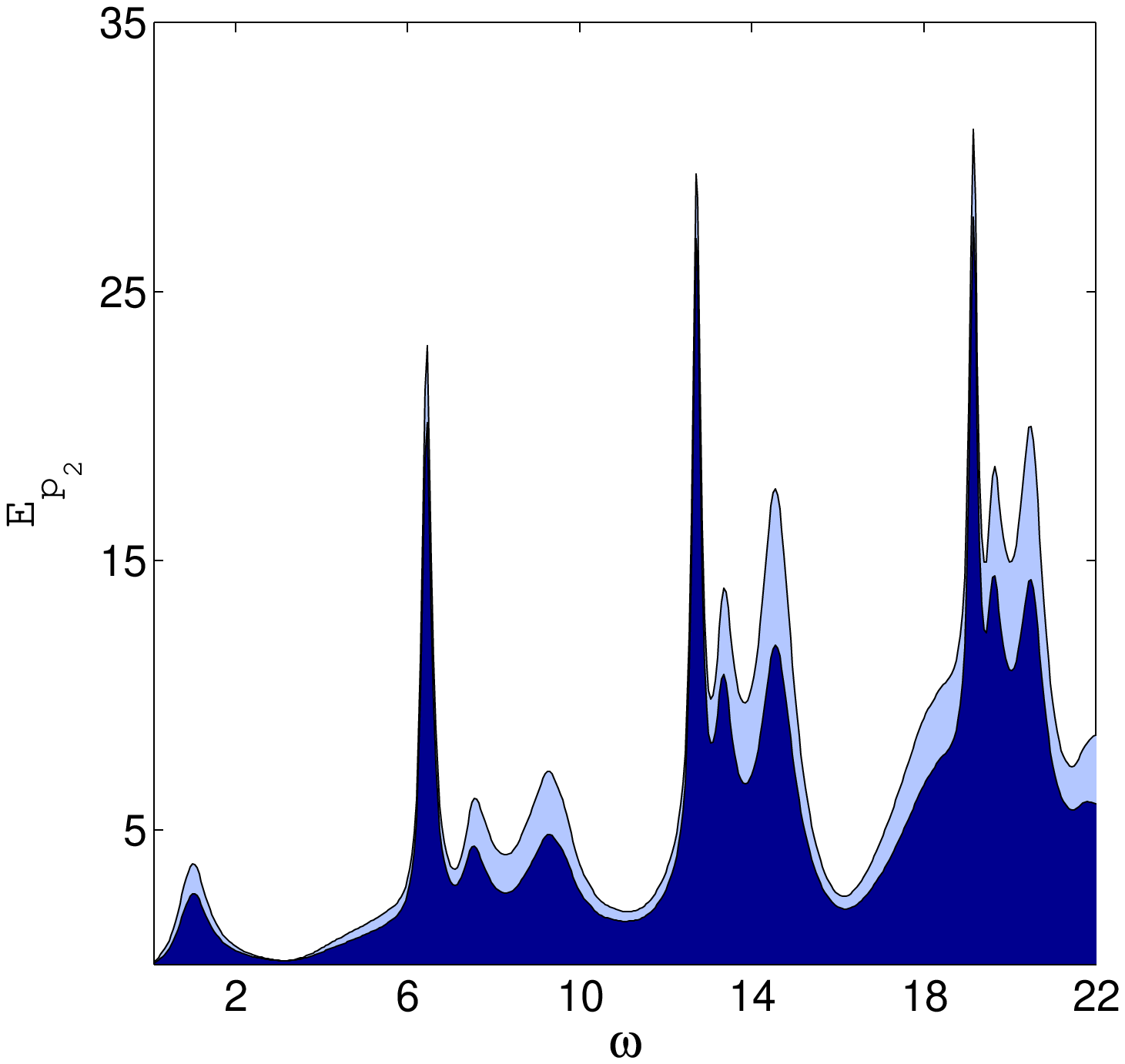}
\caption{(Colour online) Light area: the mean total energy in the left ($p_1$) and right ($p_2$) plates; dark area: the energy in the direct field $E^{(d)}_{p_1} $. The inset in the left figure shows the ratio of the energy in the direct field $E^{(d)}_{p_1} $ to the total energy   $ E_{p_1}$} \label{Fig5}
\end{center}
\end{figure}

In Fig.\ \ref{Fig4}, we present results for the high damping case. The response was found as an average over ten realizations of the random boundaries. One can see a good agreement between the hybrid result and the full FEM calculation in both stochastic subsystems. Note the difference in scale for subsystem $p_1$ and $p_2$; as $p_2$ is next to the deterministic subsystem containing the source and damping is relatively large, most of the energy is stored in this subsystem. In addition, more than 80\% of the energy is stored in the direct field over the whole frequency range, see Fig.\  \ref{Fig5}. The response of each individual  realisation of the ensemble thus coincides well with the ensemble average as shown in Fig.\ \ref{Fig4}.  The signal in $p_1$ , on the other hand, fluctuates wildly over the different ensemble realisations. The hybrid method yields the ensemble averaged response and the amount of energy in the direct and reverberant field components are comparable. In Fig.\ \ref{Fig5}, we plot the direct field energies $E^{(d)}_{p_1}$, $E^{(d)}_{p_2}$ in regions $p_1$ and $p_2$ together with the total energies $\langle E_{p_1} \rangle$,  $\langle E_{p_2} \rangle$ as function of the frequency. The resonance features of the response in Fig.\ \ref{Fig5} are mostly due to resonances in the subsystem $n_2$.  Note that the resonance positions in both field components roughly coincide, although small shifts are observable. The relative intensity of resonance peaks is also different for the two field components. Clearly, the reverberant and direct field couple differently with the resonances in the deterministic subsystems.

\begin{figure} [t] 
\begin{center}
\includegraphics[width=0.45\textwidth, height=0.45\textwidth]{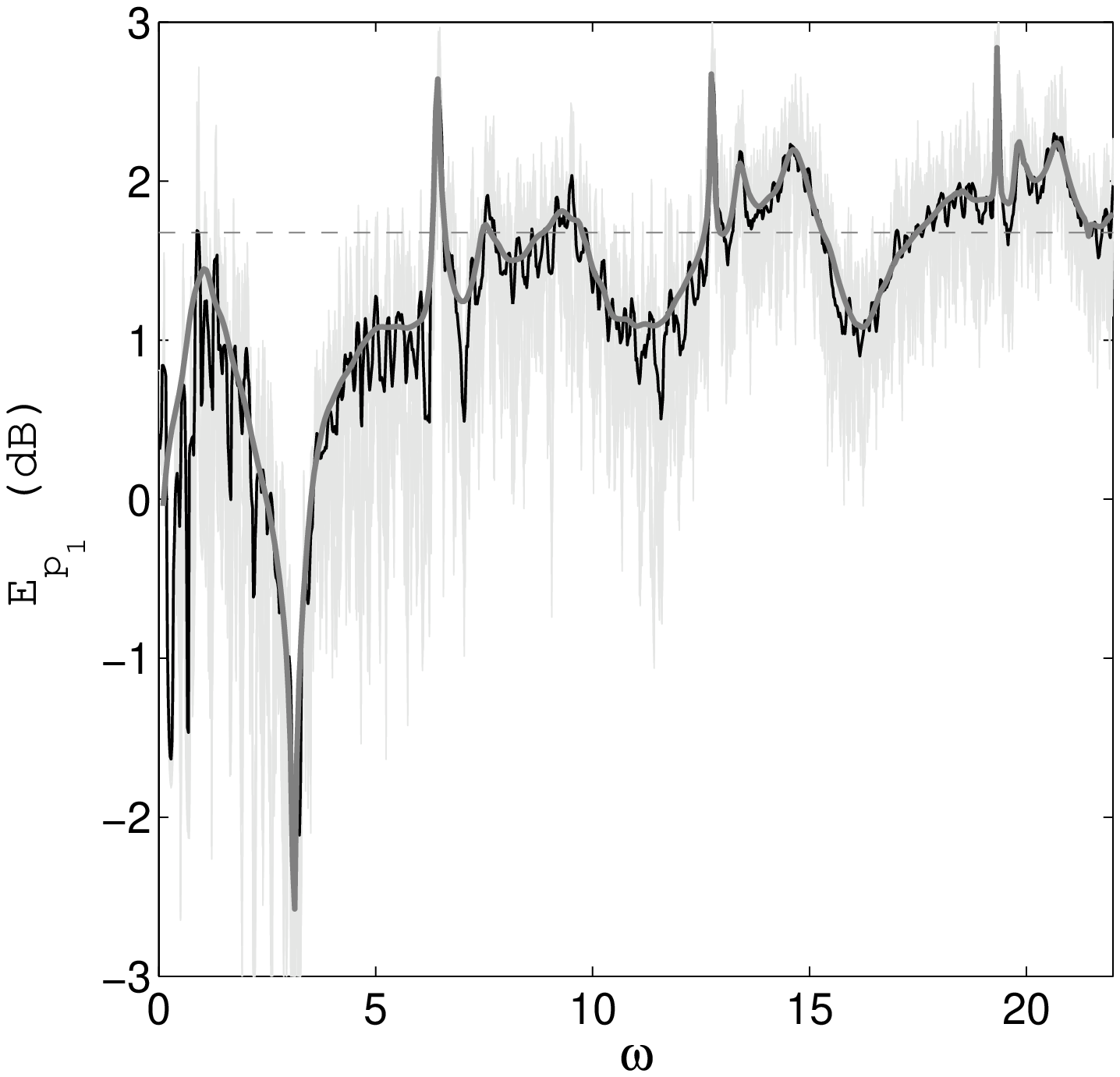}
\includegraphics[width=0.45\textwidth, height=0.45\textwidth]{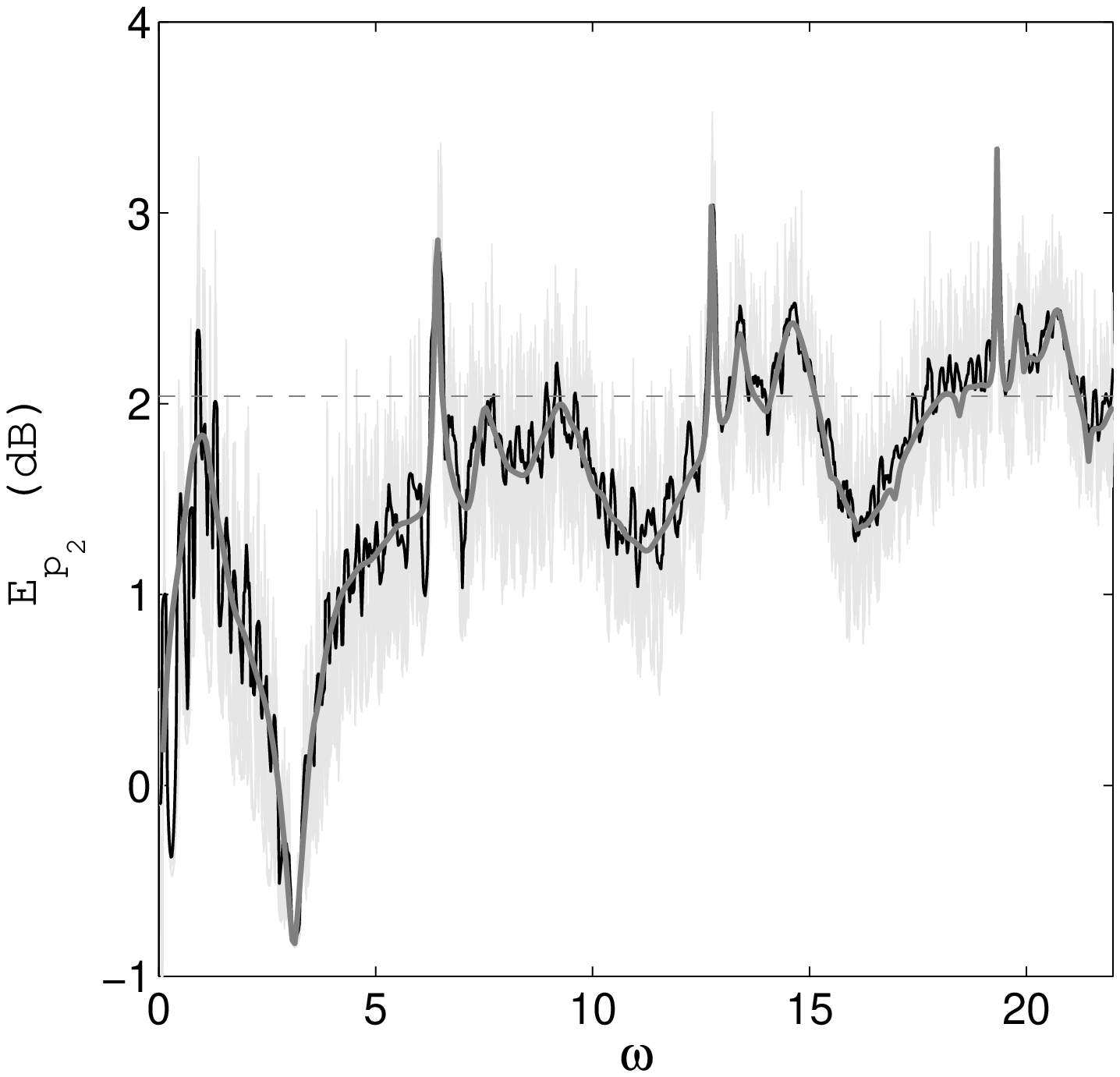}
\caption{ The low damping regime: energies in the left and right region found with Monte Carlo sampling compared to the results obtained with the hybrid method. Light grey: each Monte Carlo realization; black: ensemble average; bold grey: hybrid method; dashed line the SEA result; $\eta=0.01$.} \label{Fig6}
\end{center}
\end{figure}

In the high damping regime, the frequency dependence of the response in the FEM calculations is relatively smooth because of the high modal overlap, see Fig.\ \ref{Fig4}. That is, individual resonances in the stochastic subsystems tend to be broad and are not resolved. In contrast, in the low damping regime, one observes sharp peaks in each individual FEM realization which are associated with individual resonances of the stochastic subsystems. As a consequence, the response function varies considerably across the ensemble and individual realizations have little in common apart from the resonance structure originating from the deterministic subsystems. The computational results are presented in Fig.\ \ref{Fig6} where one can clearly see the response variation across the ensemble. (Note that results are shown on a logarithmic scale). The response was found as an average over 36 realizations. Again, one observes a good agreement between the hybrid method and the ensemble average. In Fig.\ \ref{Fig6},
we also present the results for the mean energy obtained by conventional SEA. One can see that SEA is unable to resolve any features due to resonant properties of the deterministic subsystems. It just gives a ballpark figure for the expected wave intensities which is a constant for our damping model (with damping parameter $\eta$ independent of $\omega$). 


\section{Conclusion}

A hybrid FEM/SEA method has been developed which works efficiently in the mid-frequency regime. The key features of the method are an exact computation of the direct field over the whole system and the use of random diffuse field correlation functions to evaluate the coupling constants between reverberant field components corresponding to different stochastic subsystems. The direct and reverberant fields are then coupled through energy balance equations after FEM models for both the reverberant and the direct field have been solved. The FEM model for the direct field acts globally across the  system coupling local deterministic models via direct field propagators. For the stochastic components, we solve the wave problem locally by finding Green functions for the deterministic subsystem.
Our method has been tested in both the high and low damping regime. For high damping, a large part of the energy is stored and dissipated in the direct field. The energy distribution in the direct field is highly non-uniform within each stochastic subsystem and coupling between different stochastic subsystems can therefore not be predicted by SEA \cite{Tanner}. The method developed here makes it possible to overcome this difficulty and introduces coherent coupling mechanism between different deterministic subsystems. It is valid for all penetration depths of acoustic vibrations into the medium and is able to describe the cross-over from the high to the low damping regime when most of energy is stored in the reverberant field. It, furthermore, resolves resonance structures due to the deterministic subsystems in both the direct and reverberant field.

Like the hybrid method developed in \cite{Shorter, Cotoni}, we determine local coupling parameters between stochastic subsystems using FEM which takes into account excitations through a diffuse field in the stochastic subsystems. In contrast to  Shorter and Langley, our approach makes directly use of the result Eq.\ (\ref{99}) for the field-field correlation function instead of using a reciprocity relationships between direct field radiation and diffuse reverberant loading \cite{Shorter1}. In addition, we calculate the direct field explicitly which is particularly important in the high damping limit. Thus, the method is applicable for over-damped structures where pure SEA may not produce a reliable result \cite{leBot}.

\section*{Acknowledgments}

We thank David Chappell and Niels S\o ndergaard for stimulating discussions and reading the manuscript. Fruitful discussions with Brian Mace, Frank Vogel and Cathleen Seidel are acknowledged.  We have received financial support through the European Community's Seventh Framework Programme (FP7/2007-2013) under grant agreement PIAP-GA-2008-230597.

\end{document}